\documentclass[journal,letterpaper]{IEEEtran_nssmic}
\usepackage{cite}      % Written by Donald Arseneau
\usepackage{graphicx}  % Written by David Carlisle and Sebastian Rahtz
\begin{document}
\title{Performance of the TPC with Micro Pixel Chamber Readout: micro-TPC}
\author{Kentaro~Miuchi, Hidetoshi~Kubo, Tsutomu~Nagayoshi, Atsuhiko~Ochi, Reiko~Orito, Atsushi~Takada, Toru~Tanimori, Masaru~Ueno\thanks{
Manuscript received on December 1st, 2002.}\thanks{
K.~Miuchi, H.~Kubo, T.~Nagayoshi, R.~Orito, A.~Takada, T.~Tanimori, and M.~Ueno are with the Department of Physics, Graduate School of Science, Kyoto University, Sakyo-ku, Kyoto 606-8502, Japan.}\thanks{A.~Ochi is with the Department of Physics, Graduate School of Science and Technology, Kobe University, Kobe 657-8501, Japan}}
%\thanks{Manuscript received January 20, 2002; revised August 13, 2002.
%\markboth{Journal of \LaTeX\ Class Files,~Vol.~1, No.~8,~August~2002}{Shell \MakeLowercase{\textit{et al.}}: Bare Demo of IEEEtran.cls for Journals}

% If you want to put a publisher's ID mark on the page
% (can leave text blank if you just want to see how the
% text height on the first page will be reduced by IEEE)
%\pubid{0000--0000/00\$00.00~\copyright~2002 IEEE}

%\specialpapernotice{(Invited Paper)}

\maketitle

\begin{abstract}
Micro-TPC, a time projection chamber (TPC) with micro pixel chamber ($\mu$-PIC) readout was developed for the detection of the three-dimensional fine ($\sim$ sub-millimeter) tracks of charged particles. We developed a two-dimensional position sensitive gaseous detector, or the $\mu$-PIC,  with the detection area of 10 $\times$ 10 cm${}^{2}$ and 65536 anode electrodes of 400 $\mu$m pitch. We achieved the gas gain of more than $\rm 10^4$ without any other multipliers. With the pipe-line readout system specially developed for the $\mu$-PIC, we detected X-rays at the maximum rate of 7.7~MHz. We developed a micro-TPC with the $\mu$-PIC and three-dimensional tracks of electrons were detected with the micro-TPC. We also developed a prototype of the MeV gamma-ray imaging detector which is a hybrid of the micro-TPC and NaI (Tl) scintillator so that we showed that this is a promising method for the MeV gamma-ray imaging.
\end{abstract}


\section{Introduction}
\PARstart{R}{ealization} of a micro-TPC, a time projection chamber (TPC), which can detect three-dimensional fine tracks of charged particles, has been wanted for years. We developed the $\mu$-PIC, a two-dimensional fine position detector, and a fast readout electronics in order to realize such an ``electric cloud chamber''. In this paper, the performance of the micro-TPC, together with one example of its applications, is described.

%We developed $\mu$-PIC  as the two-dimensional fine position detector of the micro-TPC, which is described in Section \ref{section_uPIC}. We also developed a pipe-line readout electronics of the fine timing measurement of the micro-TPC as described in Section \ref{section_electronics}. A micro-TPC with a detection area of 10 $\times$ 10$\rm cm^2$ and a drift length of 8cm was developed and its performance was tested. We developed the prototype MeV gamma-ray imaging detector which consists of the micro-TPC and the NaI (Tl) scintillator and tested the concept of the MeV gamma-ray imaging detector. The design of the micro-TPC and the performances are described in Section \ref{section_design}  and Section \ref{section_performance}, respectively.

% which is one of the most promising applications of the micro-TPC.

%, which measures the drift times and the two-dimensional positions of the electrons from the primary ionizations\cite{TPC:Nygren}. 

% The development and the performance of the $\mu$-PIC are described in Section \ref{section_uPIC}, the development and the performance of the readout electronics are described in Section \ref{section_uPIC}
 

%\hfill November 10, 2002

\section{$\mu$-PIC DETECTOR}
\label{section_uPIC}
The Micro Pixel Chamber, or the $\mu$-PIC, is a pixel type gaseous two-dimensional imaging detector which takes over the outstanding properties of the MicroStrip Gas Chamber (MSGC\cite{MSGC:Oed, MSGC:Tanimori}) such as the good position resolution and operating capacities under high flux irradiation. The $\mu$-PIC is manufactured by the print circuit board (PCB) technology in contrast to an older concept pixel type detector, the microdot chamber\cite{Microdot}, which is made in the MOS technology. With the PCB technology, large area detectors can be made cheaply, which is an important feature for developing various kinds of applications.  As a consequence of the geometrical properties, discharge problems are less disturbing with $\mu$-PIC \cite{uPIC:Ochi1, uPIC:Ochi2, uPIC_PSD:Nagayoshi}, so that stable operations at high gas gain can be realized.

\begin{figure}[h]
\centering
\includegraphics[width=3.in]{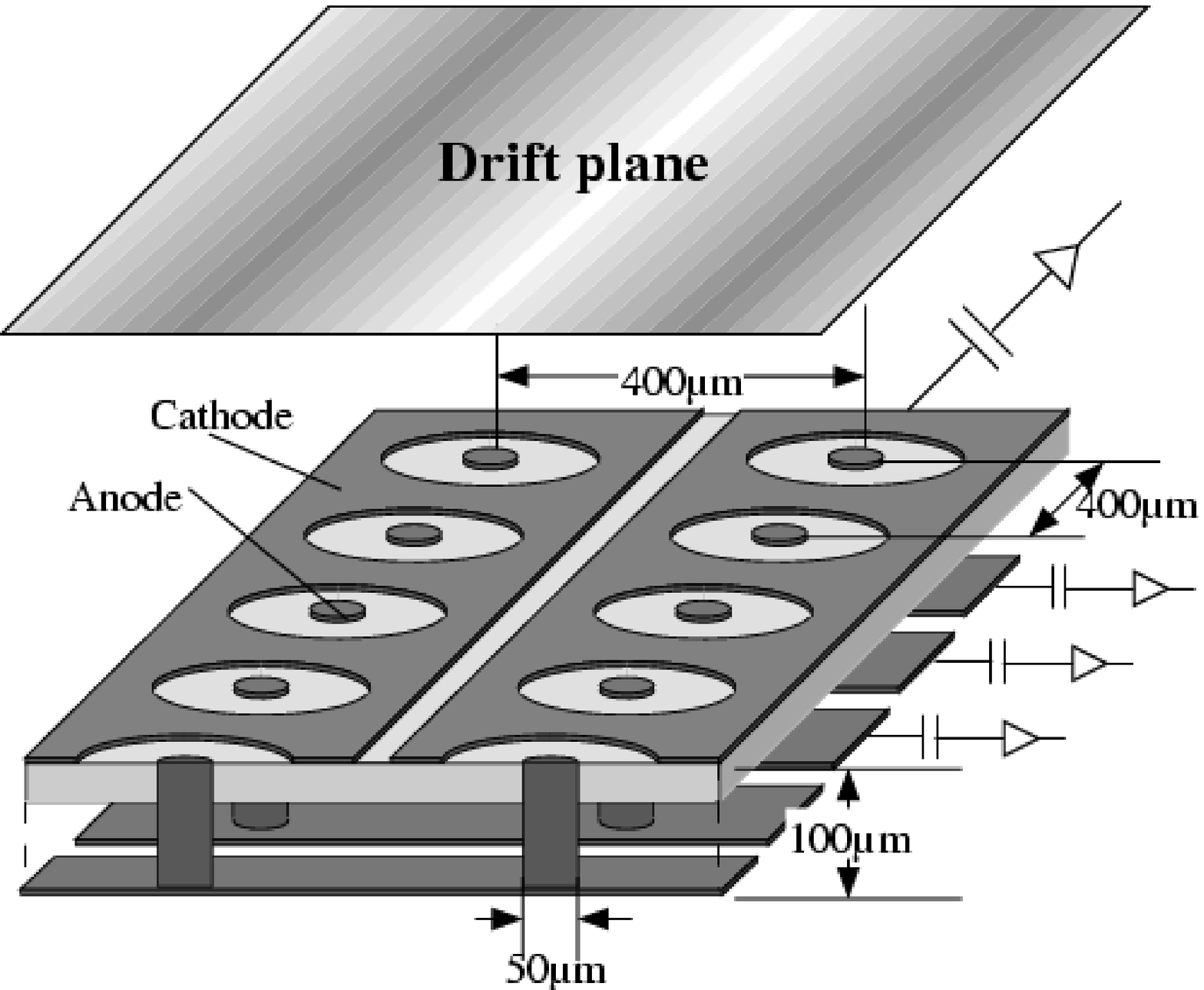}
\caption{Schematic structure of the $\mu$-PIC.}
\label{uPICstruct}
\end{figure}
The schematic structure of the $\mu$-PIC is shown in Fig.~\ref{uPICstruct}. The $\mu$-PIC is a double-sided PCB with a 100~$\mu$m-thick polyimide substrate. 256 anode strips are formed on one side of the $\mu$-PIC while 256 cathode strips are orthogonally placed on the other side. Both anode and cathode strips are placed with a pitch of 400 $\mu$m. Cathode strips have holes of 200 $\mu$m diameter and  anode electrodes of 50$\sim$70 $\mu$m diameter are formed on the anode strips at the center of each cathode hole.
% The top of the anode electrodes are 0 $\sim$ 20 $\mu$m below the surface of the polyimide substrate. 
The signals from anode strips and cathode strips are of the same size, in contrast to the MSGCs, whose pulse heights from back strips are 20$\sim$30$\%$ of those from anode strips\cite{MSGC:Tanimori}. 
We developed the $\mu$-PIC of 10 $\times$ 10~$\rm cm^2$ detection area with 256 $\times$ 256 anode electrodes. The $\mu$-PIC is mounted on the mother board of  30 $\times$ 30~$\rm cm^2$ area by the bonding technique.
%\begin{figure}
%\centering
%\includegraphics[width=2.5in]{uPICall.eps}
%\caption{Picture of the  $\mu$-PIC mother board with 10 $\times$ 10 $\rm cm^2$ detection area.}
%\label{uPIC_photo}
%\end{figure}

\begin{figure}[h]
\centering
\includegraphics[width=3.0in,height=2.5in]{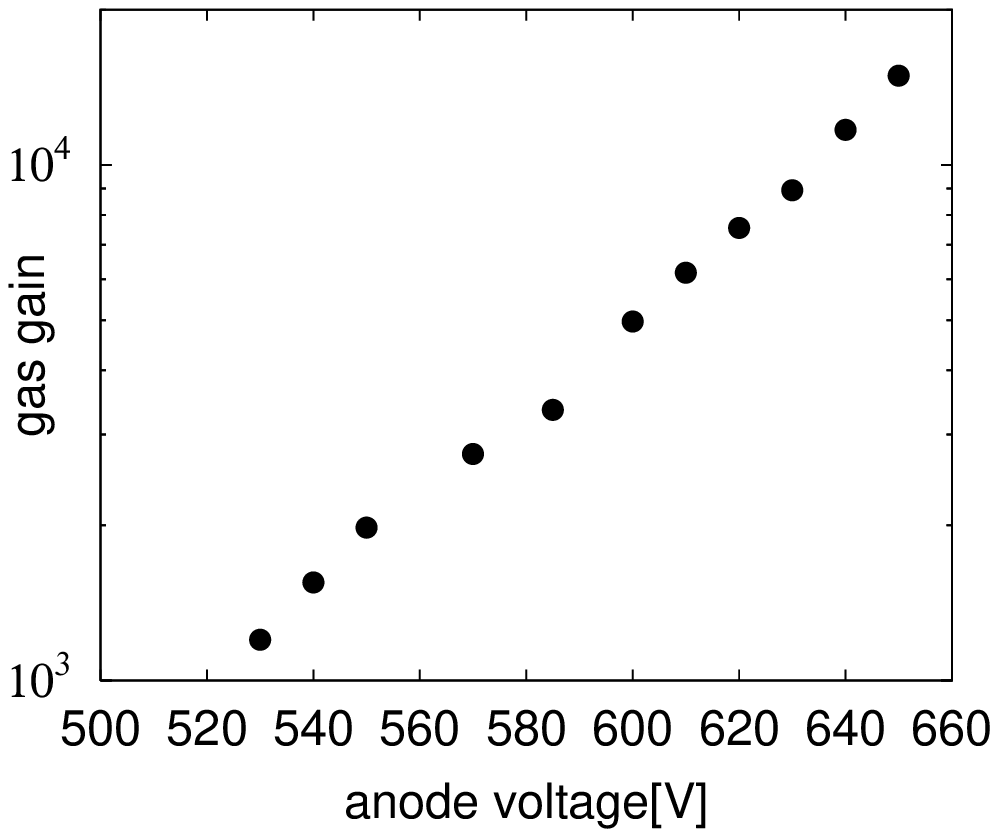}
\caption{Gas gains as a function of the anode voltage.}
\label{gain_HV}
\end{figure}

\begin{figure}[h]
\centering
\includegraphics[width=3.0in,height=2.5in]{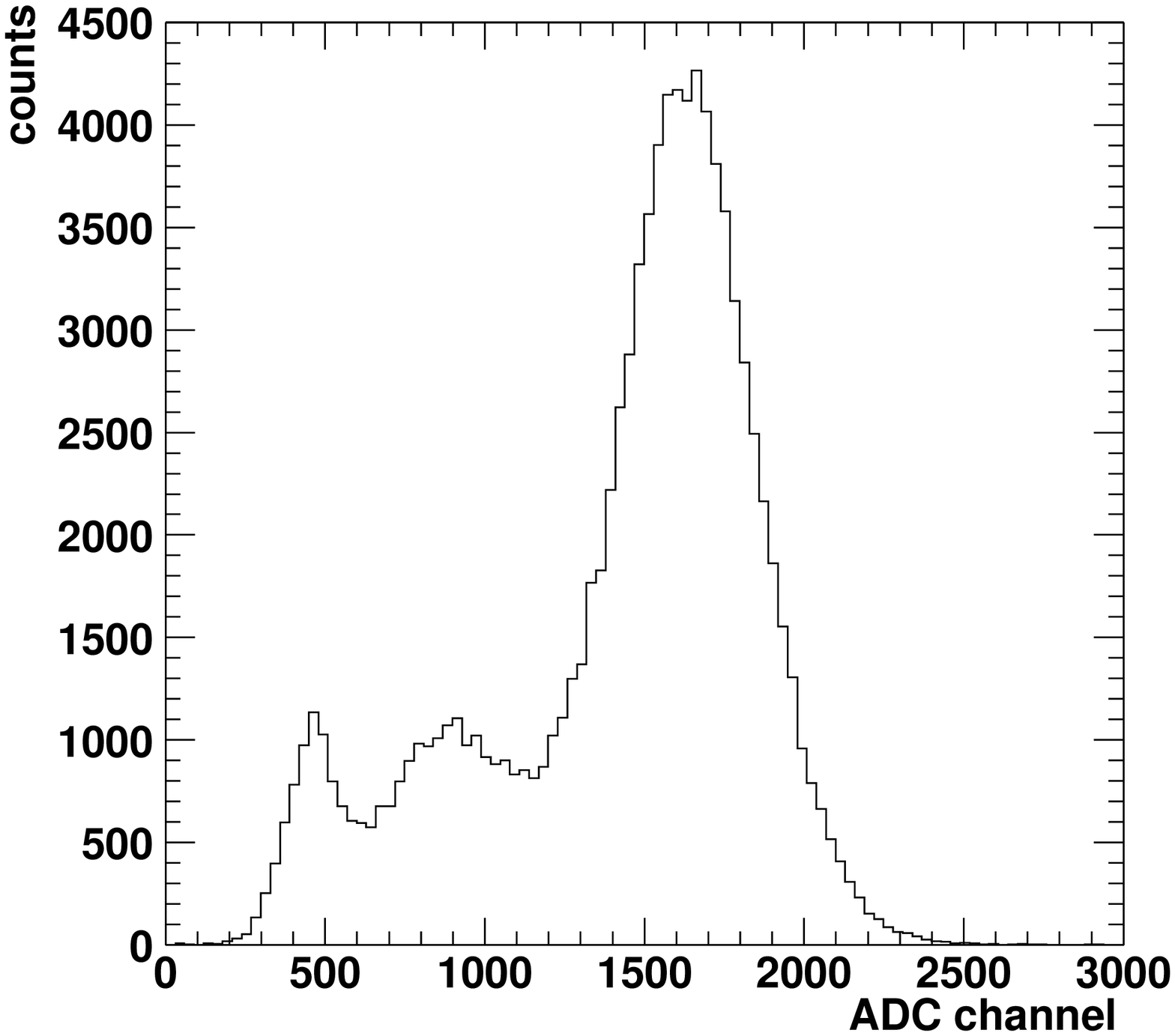}
\caption{Obtained spectrum of the X-rays from a ${}^{55}$Fe radioactive source.}
\label{Fe55_spec}
\end{figure}
We irradiated a non-collimated $\rm{}^{55}Fe$ radioactive source and measured the output charge from $32\times256$ pixels so that we calculated the gas gain. The gas mixture of argon 80$\%$ and ethane 20$\%$ at 1 atm was flowed for the measurement.  The gas gain of the $\mu$-PIC is plotted as a function of the anode voltage in Fig.~\ref{gain_HV}. We achieved a maximum gas gain of 1.5 $\times 10^4$ and the gas gain for the stable operation of 5000 without any other multipliers.
In Fig.~\ref{Fe55_spec}, the spectrum of the X-rays from the $\rm{}^{55}Fe$ radioactive source is shown. The energy resolution is $30\%$ (FWHM) for the 5.9~keV X-rays.  This result is worse than our previous results $20\%$ obtained with  the trial piece which has 3 $\times$ 3 $\rm cm^2$ detection area\cite{uPIC:Ochi1}. The non-uniformity of the anode electrodes in the large detection area makes the resolution worse. New technologies for forming the uniform electrodes are now sought for so as to achieve higher gas gain and better energy resolution.

\begin{figure}
\centering
\includegraphics[width=3.5in,height=2.in]{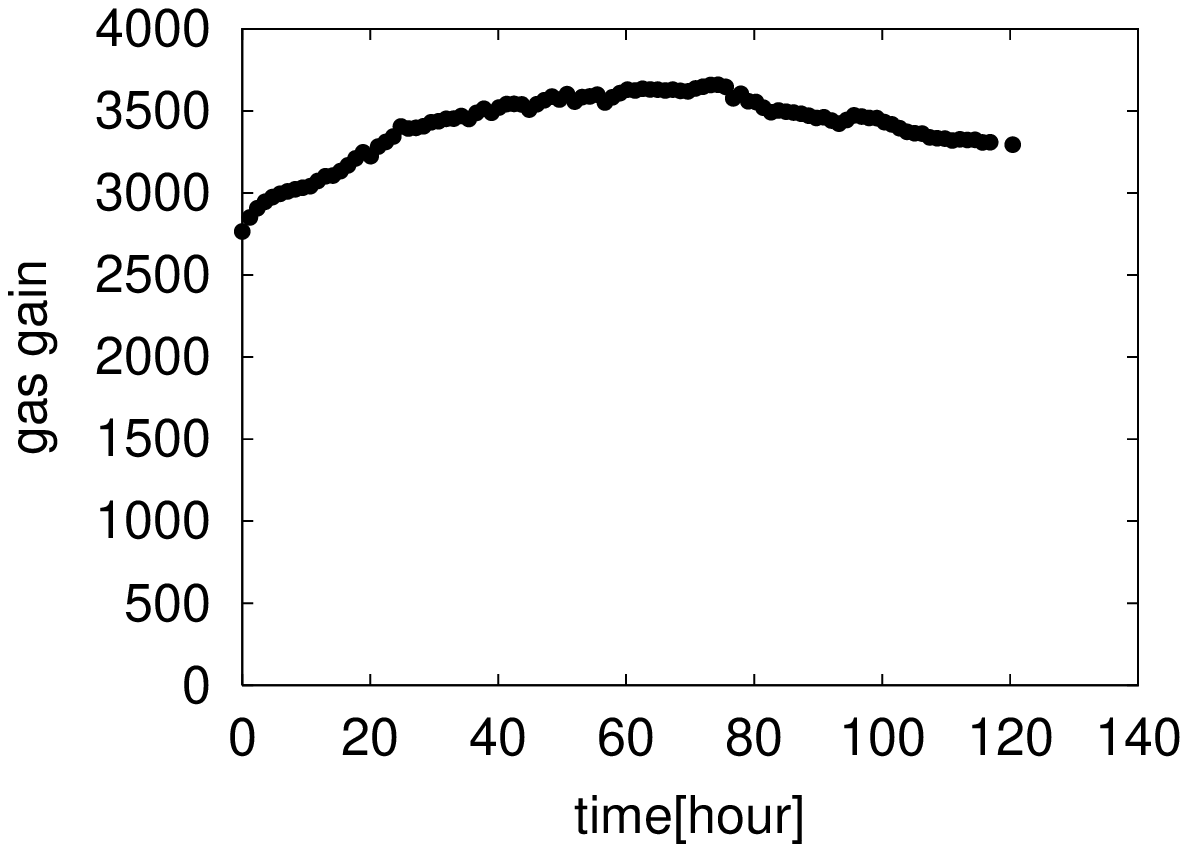}
\caption{Gas gains as a function of the time.}
\label{gain_time}
\end{figure}

Long term stability is a necessary feature in order to realize the various applications of the $\mu$-PIC. We operated the $\mu$-PIC for 120 hours with the gain of more than 3000 without serious discharge. Gas gain was measured every one hour by the same method described above. The gain variation is shown in Fig.~\ref{gain_time} as a function of the elapsed time since the anode voltage supply. Slight increase of the gain was observed in the first 20 hours, which we think is due to the polarization effect of the substrate. The gain was stable within 10$\%$ after 20 hours.

%\begin{figure}
%\centering
%\includegraphics[width=3.5in,height=2.in]{knife.eps}
%\caption{Result of the knife edge test. The shadow of the lead block is seen in the right.}
%\label{knife}
%\end{figure}

\begin{figure}
\centering
\includegraphics[width=3.5in]{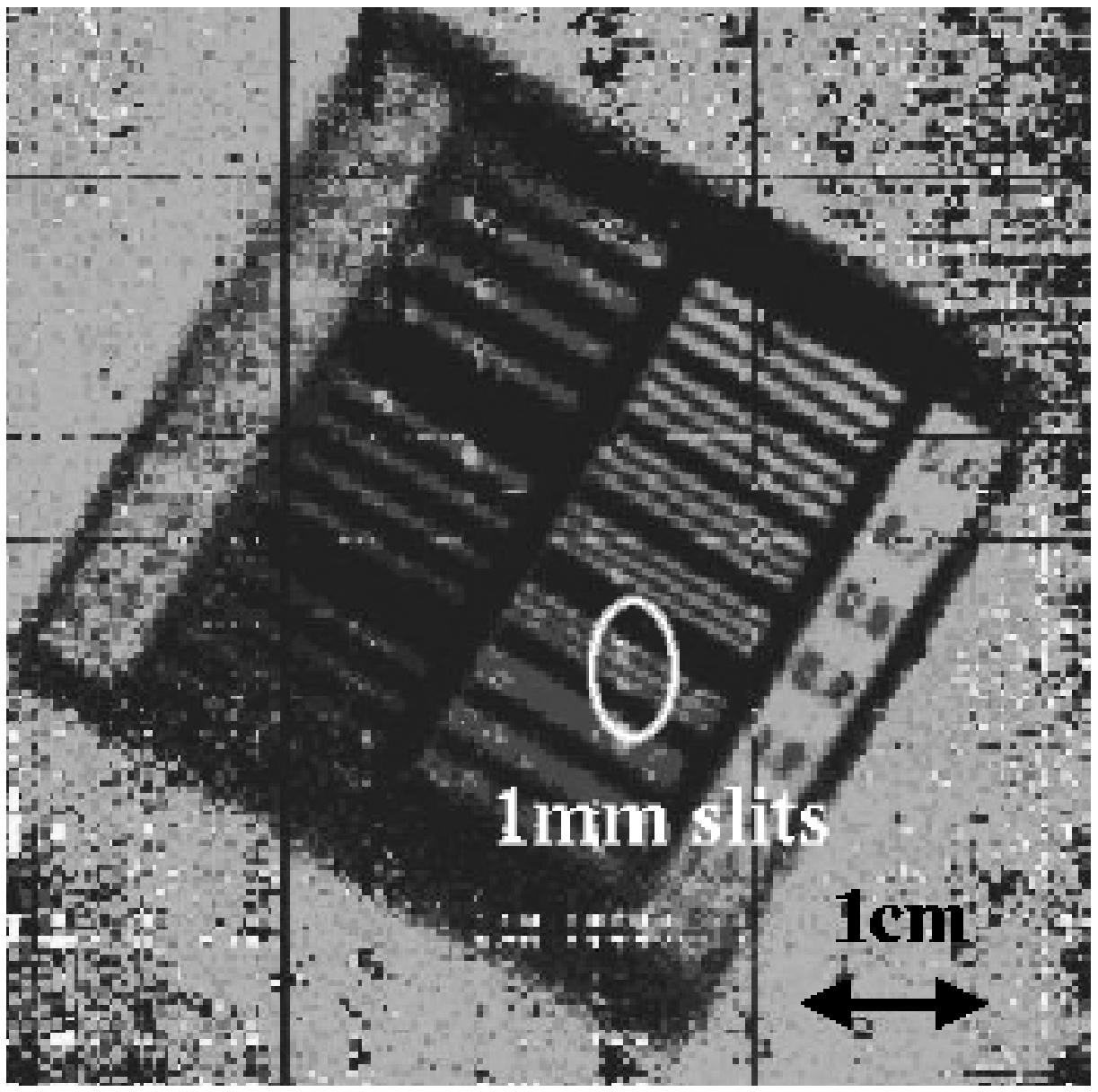}
\caption{The X-ray image of the test chart.}
\label{test_image}
\end{figure}

We took the X-ray image of the test chart with an X-ray generator (Kevex X-Ray CU028, tungsten target). The acceleration voltage of the X-ray generator was 18~kV. Low energy X-rays were filtered with 50~$\mu$m thick copper and 1 mm thick aluminum and the energy of the X-rays was distributed between 14~keV and 18~keV with the intensity peak at 17~keV. We used the gas mixture of xenon 70$\%$ and ethane 30$\%$ at 1 atm. We chose xenon gas in order to achieve clear images.
% since t
The practical range of the 20~keV electron in argon gas is 3 mm while that in the xenon gas is 1 mm. 
 The detection depth was set to be 2mm.  The test chart is 5 $\times$ 5 $\rm cm^2$ area and slits of various widths are scribed in the thin lead layer on the plastic plate.  In Fig.~\ref{test_image}, obtained X-ray image is shown. The slits of 1 mm width are seen clearly.  

\begin{figure}
\centering
\includegraphics[width=3.0in]{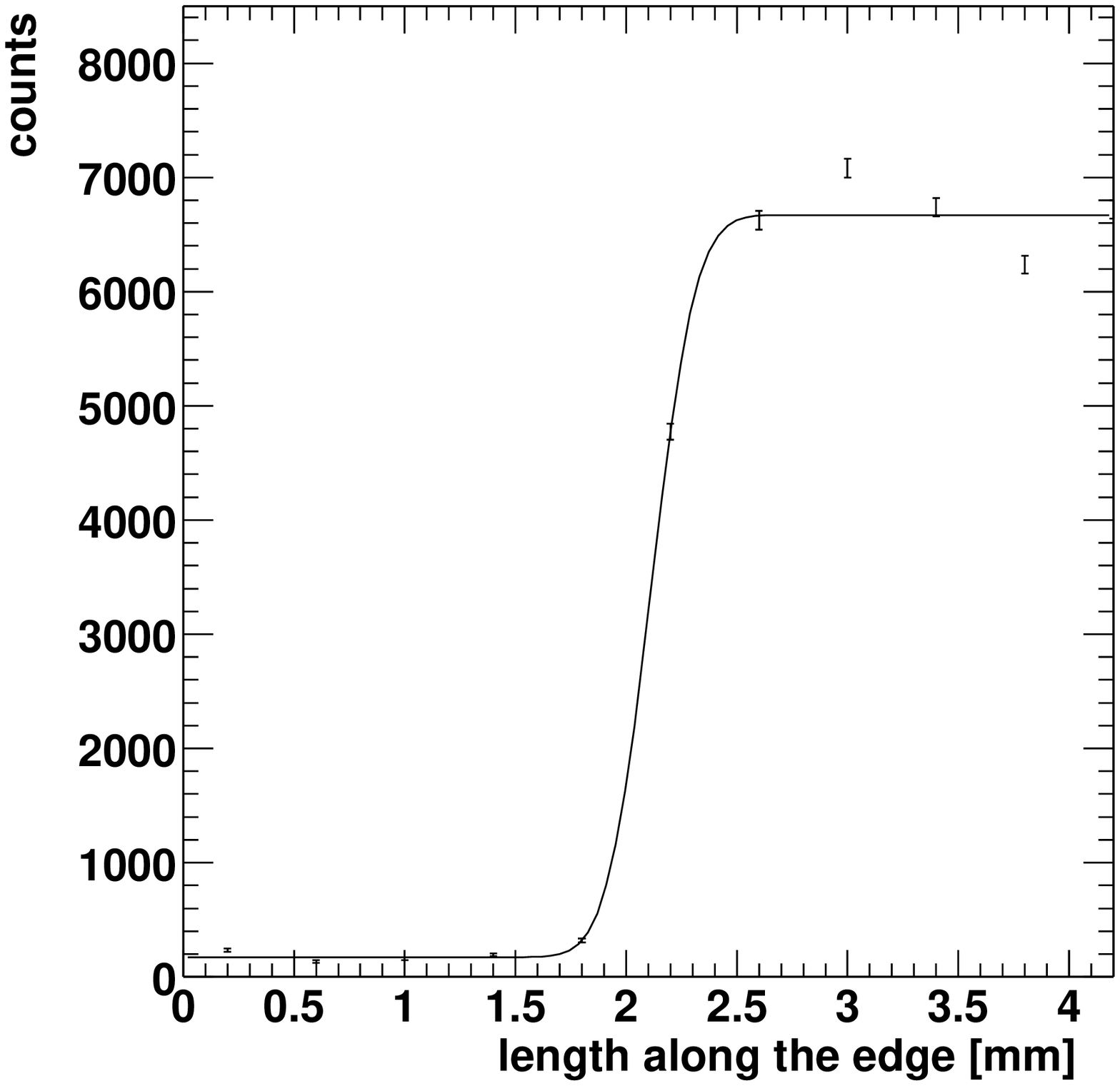}
\caption{Projected image of the test chart edge and the best fit function.}
\label{knife}
\end{figure}

We measured the spatial resolution from the edge image of the test chart. Acceleration voltage of the X-ray generator was set at 10~kV for this measurement in order to suppress the range of the electrons in the xenon gas. The active area was set to be $4\times8 cm^2$ with one of the edges at the center of the active area. The test chart image was projected along one of the edges for 1.3 cm.  The projected image was fitted with the function $a_1+a_2{\cdot}erf((x-a_3)/\sqrt{2}a_4)$, where $erf(x)=2/\sqrt{\pi} \cdot \int_{0}^{x}exp(-t^2)dt$ is the error function and $a_1, a_2, a_3$ and $a_4$ are the fitting parameters.  The calculated spatial resolution($a_4$) was 160 $\mu$m. The histogram of the projected image of the test chart and the best fit function is shown in Fig. \ref{knife}.

\section{READOUT ELECTRONICS}
\label{section_electronics}
\begin{figure}
\centering
\includegraphics[width=3.5in]{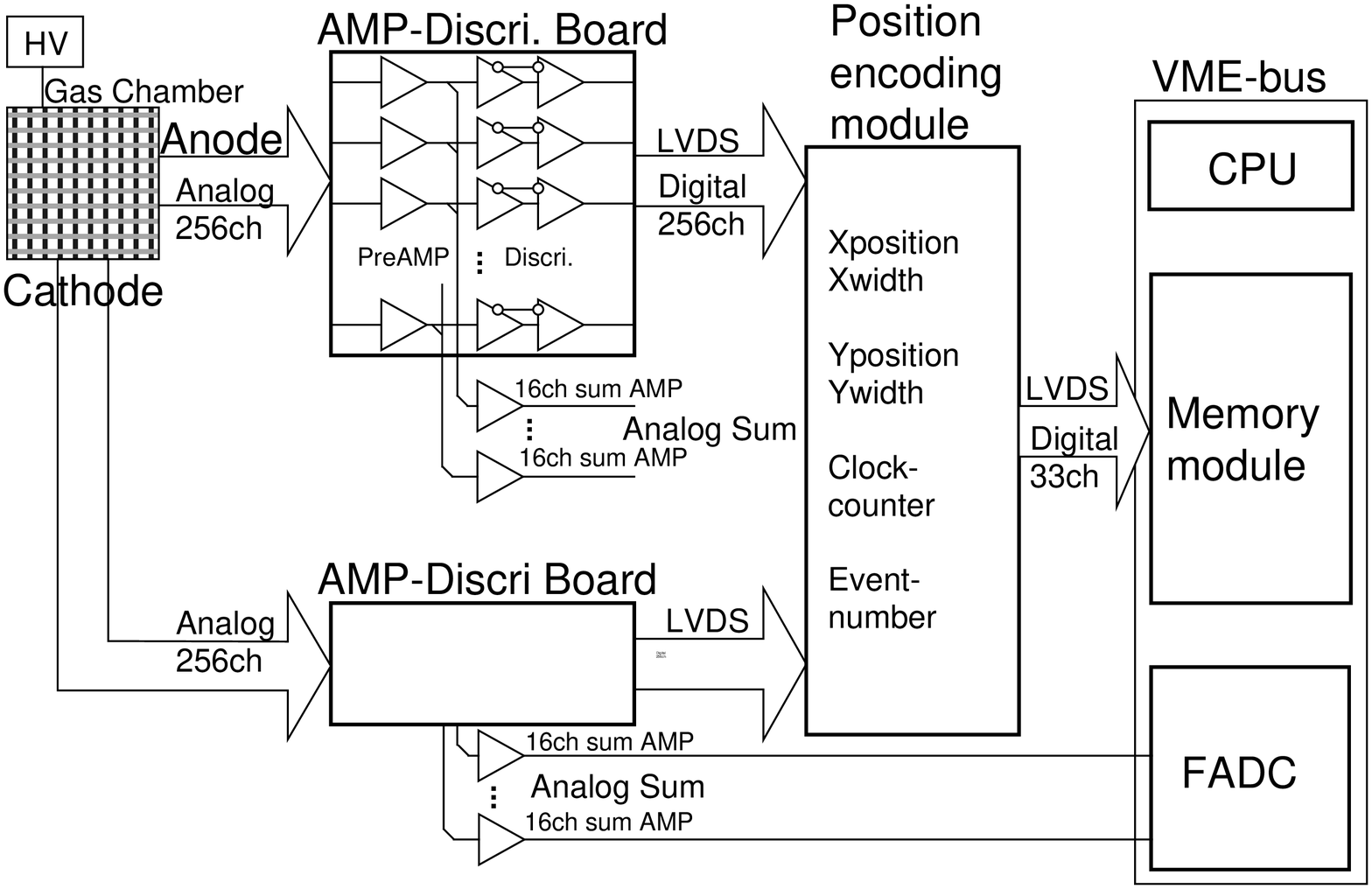}
\caption{Block diagram of the readout electronics. }
\label{DAQ_block}
\end{figure}

A pipe-line readout electronics system for the $\mu$-PIC was developed as well as the detector itself\cite{TPC_PSD:Kubo}. The block diagram of the readout electronics is shown in Fig. \ref{DAQ_block}. 
Signals from the $\mu$-PIC are amplified and discriminated by the preamplifier cards attached to the rear of the mother board. The preamplifier cards output both analog and digital signals. The digital signals are encoded by the position encoding module which consists of five Field Programmable Gate Arrays (FPGAs) and recorded by the memory module which works at the clock of 20 MHz.  Analog signals from the 16 cathode strips are summed and digitized by the 100~MHz flash ADC (FADC:REPIC RPV-160) to determine the event energy.

\begin{figure}
\centering
\includegraphics[width=3.0in]{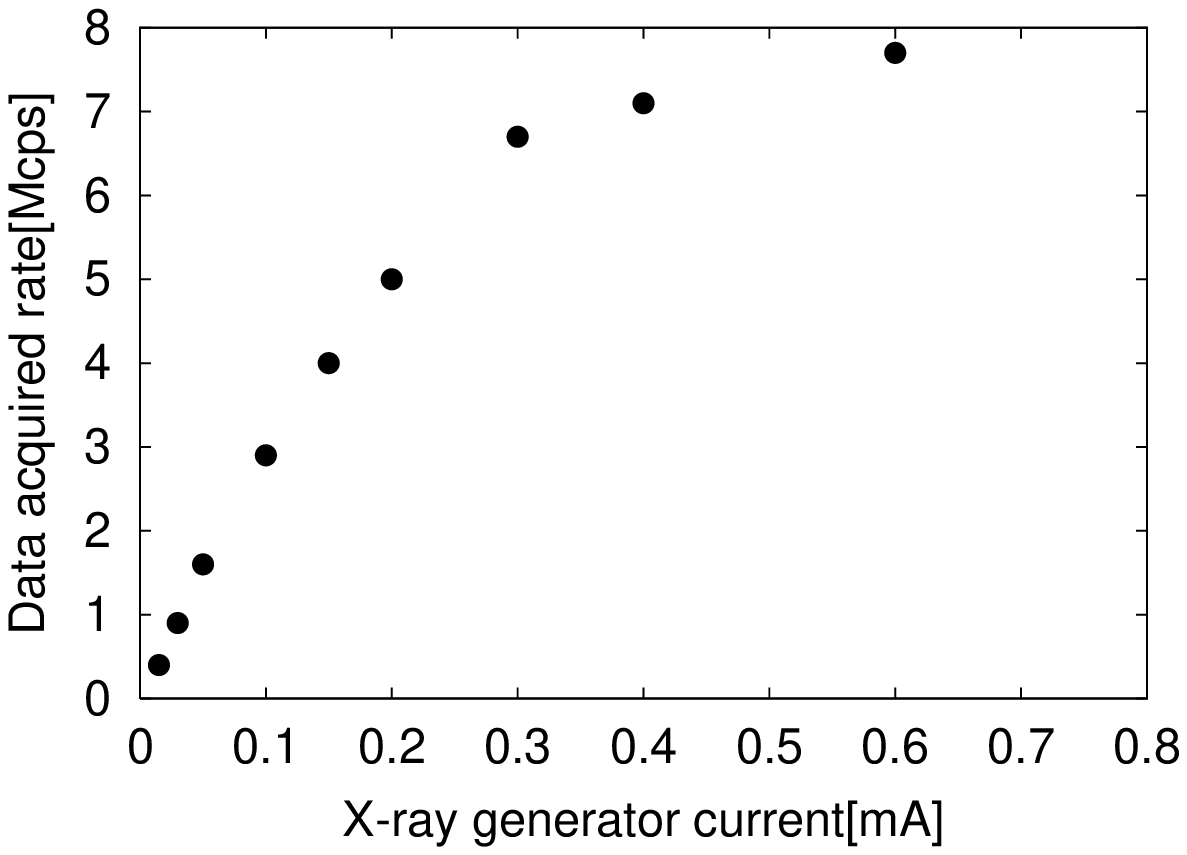}
\caption{Result of the high rate test of the $\mu$-PIC and the readout system.}
\label{highratetest}
\end{figure}

In order to check the performance of the $\mu$-PIC and the readout system for high rate signals, we irradiated the $\mu$-PIC with the X-rays from the X-ray generator. The acceleration voltage was 20 kV and the intensity was controlled by way of varying the tube current between 0.01 mA and 0.6 mA.  The data acquisition rate is plotted against the X-ray intensity in Fig.~\ref{highratetest}. We achieved the maximum data acquisition rate of 7.7 MHz. This rate is twice higher than our previous readout system\cite{MSGC:Ochi}. The saturation is thought to be due to the software for the FPGAs and optimizations are now under way.
%, which is determined by the clock of the encoding system (40 MHz) and the requirement of one and only one hit for the valid event. This rate is four times higher than out previous readout system\cite{MSGC:Ochi}.
%The results are shown in Fig.~\ref{highratetest}. The data acquisition rates are plotted against the X-ray intensities. The maximum data acquisition rate of 14MHz was achieved, which is determined by the clock of the encoding system (40 MHz) and the requirement of one and only one hit for the valid event.

\section{DESIGN OF THE MICRO-TPC}
\label{section_design}
\begin{figure}
\centering
\includegraphics[width=3.5in]{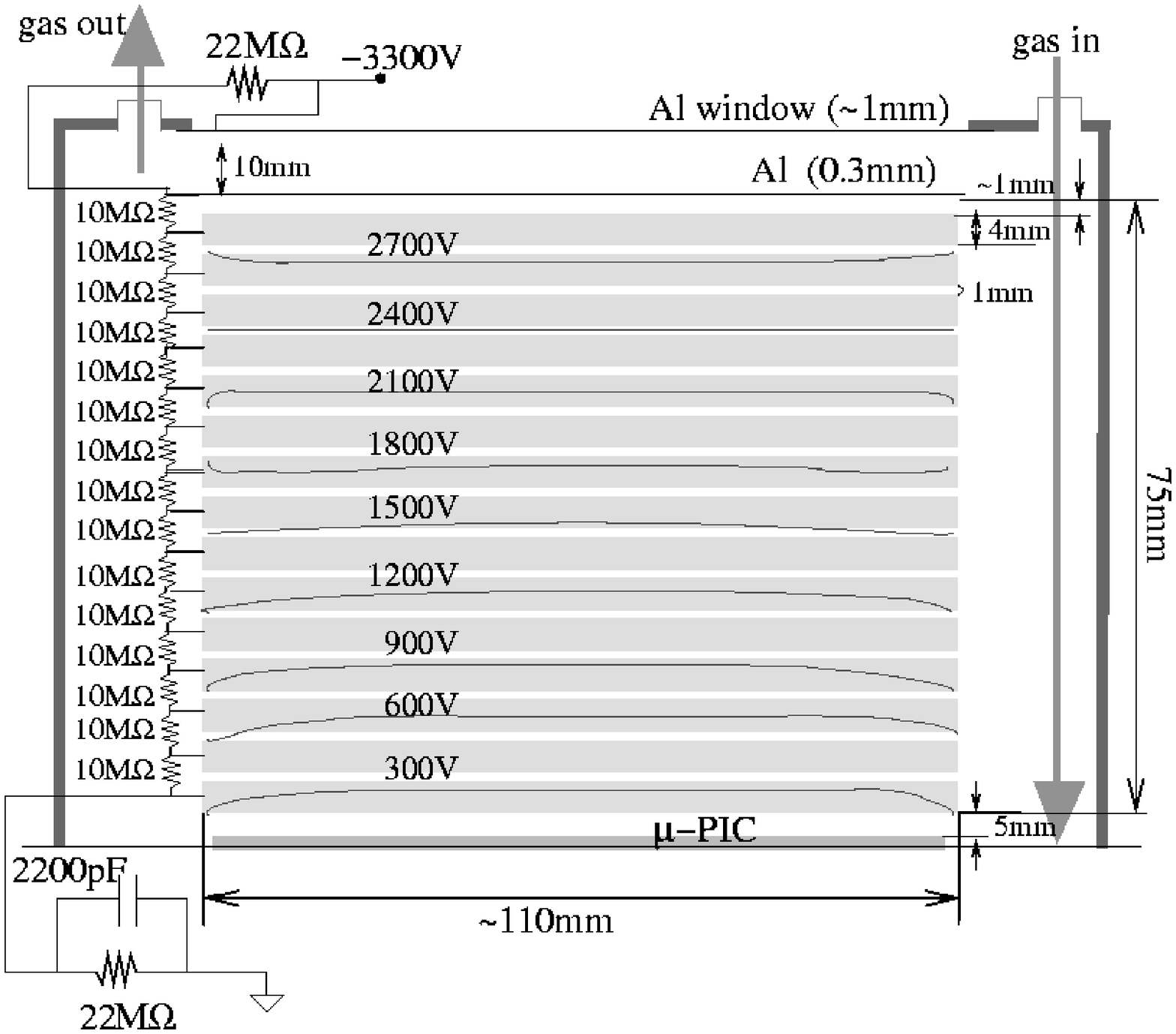}
\caption{Structure of the time projection chamber.}
\label{TPC_structure}
\end{figure}

With the $\mu$-PIC and the readout electronics system described above, we developed a time projection chamber (TPC) which has fine time and spatial resolutions\cite{TPC_PSD:Kubo}.  A gas enclosure with a drift length of 8 cm was attached to the $\mu$-PIC. Schematic drawings of the TPC gas enclosure is shown in Fig. \ref{TPC_structure}. A negative high voltage of 3.3 kV is supplied to the 1 mm thick aluminum window which is connected to the aluminum plane of 0.3mm thick through 22 M$\Omega$. Fifteen field cage electrodes with 4 mm widths and 1mm spacings are connected to the aluminum plane in series through 10 M$\Omega$ resistors.
% According to the calculation by Garfield 7 \cite{Garfield:Veenhof}, 
The electric field of 0.4 kV/cm is produced uniformly in the inner region.

\section{PERFORMANCE OF THE MICRO-TPC}
\label{section_performance}
%Performances of the micro-TPC was measured 

%\begin{figure}
%\centering
%\includegraphics[width=2.5in]{uPIC-TPC.ps}
%\caption{}
%\label{}
%\end{figure}
\subsection{Drift Velocity}
\begin{figure}
\centering
\includegraphics[width=3.in]{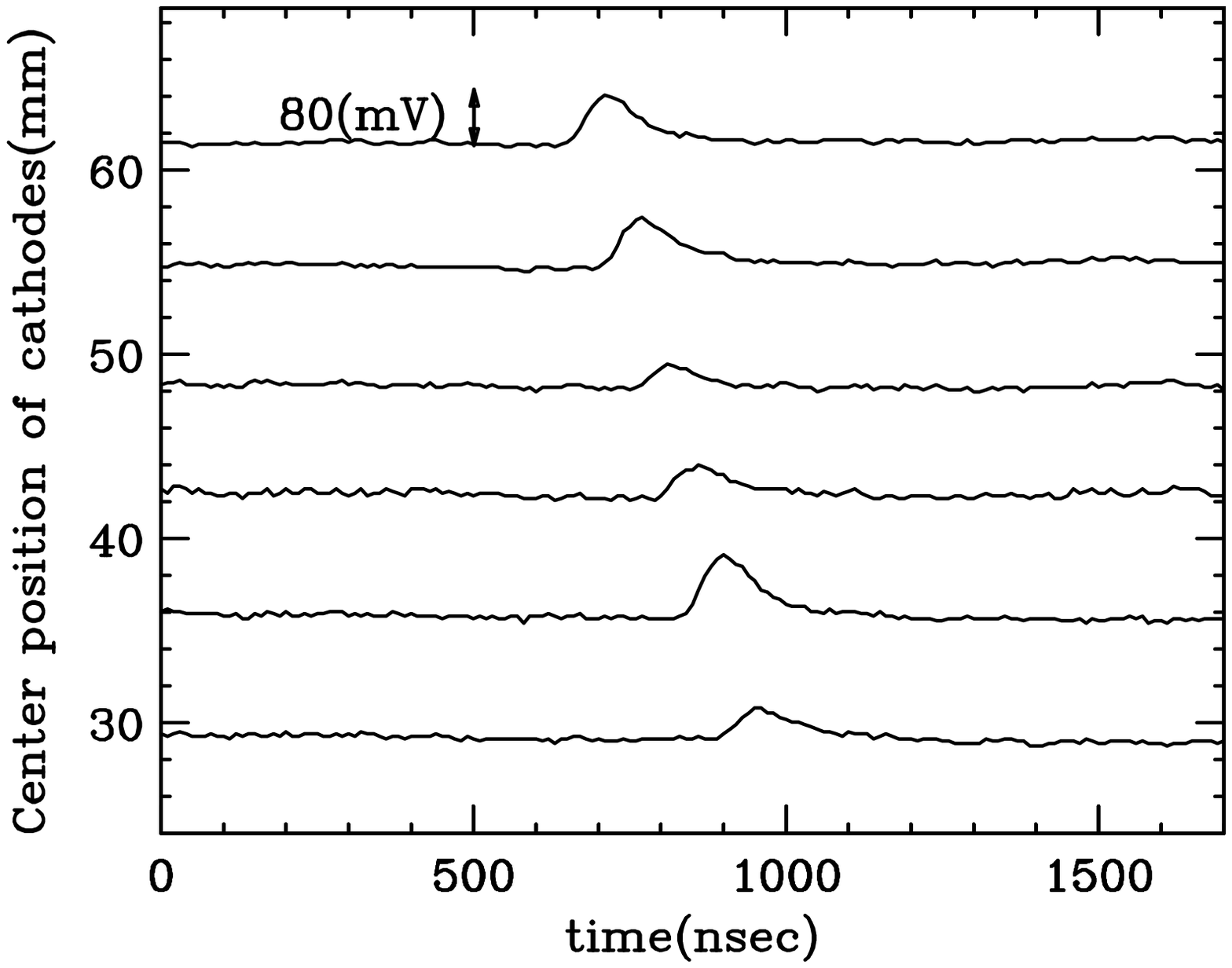}
\caption{A typical cosmic ray muon signal read from the cathode strips.}
\label{FADC_muon}
\end{figure}

\begin{figure}
\centering
\includegraphics[width=3.0in]{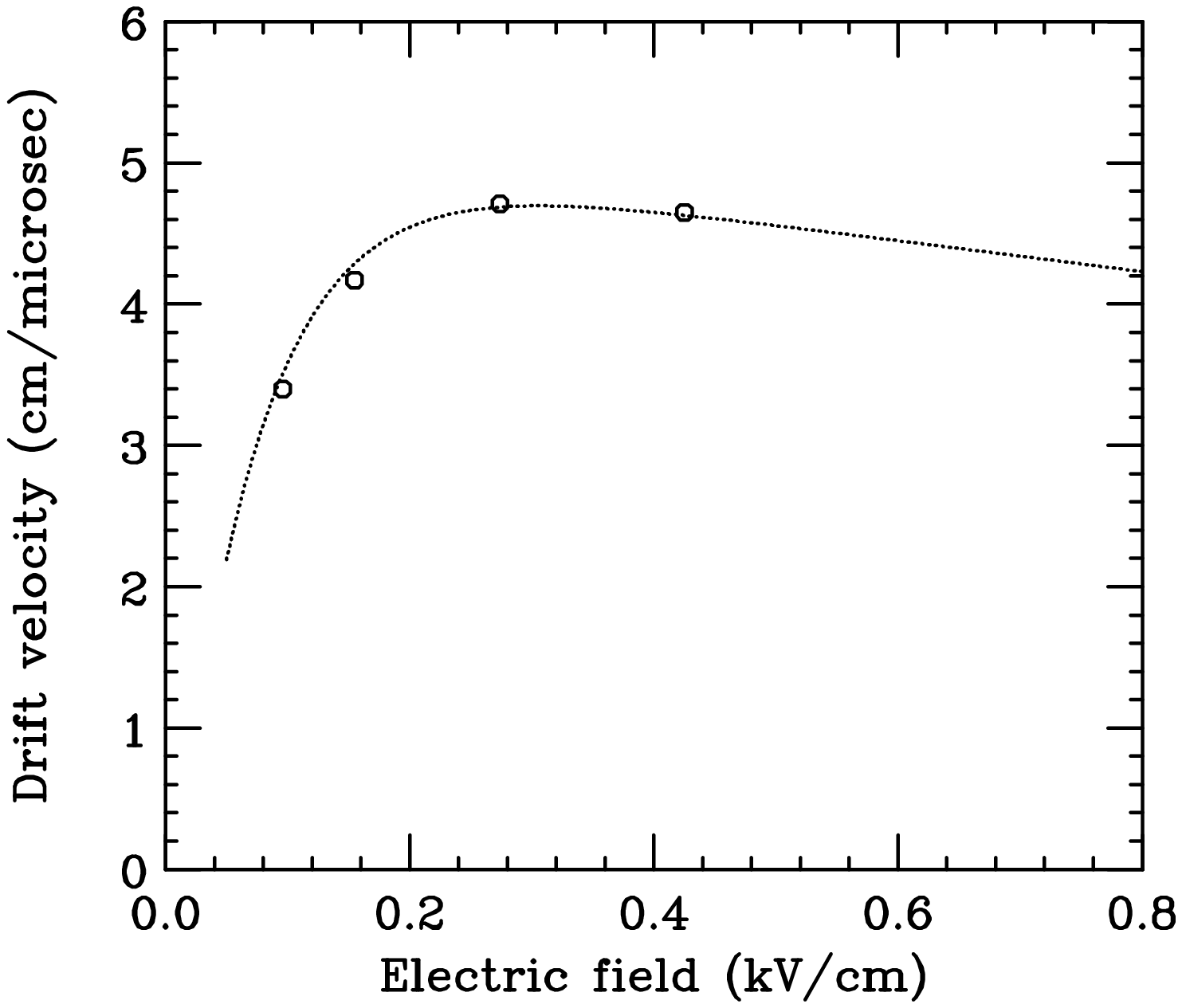}
\caption{Measured drift velocity of electrons as a function of the electric field (circle). The dotted line is the result of another group\cite{driftV:Becker}}
\label{driftV}
\end{figure}

We measured the drift velocity of electrons and the gain dependence on the drift length in the micro-TPC using the cosmic-ray muons. We placed two plastic scintillators on both sides of the micro-TPC perpendicular to the direction of the electric field. We acquired the micro-TPC events in coincidence with both of the two scintillators. 
%The gas mixture of argon 80$\%$ and ethane 20$\%$ at 1 atm was flowed. 
Fig. \ref{FADC_muon} shows the shapes of signals from cathode electrodes for a cosmic-ray muon. The signals are not identical because of the fluctuation of the gas gain due to the shape of the anode electrodes. The drift velocity was measured at various electric fields. The result shown in Fig.~\ref{driftV} is consistent with the previous measurement \cite{driftV:Becker}.  No gain difference was observed throughout the drift length.
%The gas gain was estimated to be 3000. The position resolution was 0.2mm (rms). The detected count rate was constant for three days, and the TPC worked without deterioration. 
%\begin{figure}
%centering
%\includegraphics[width=2.5in]{dat12_700evt.eps}
%\caption{}
%\label{driftV}
%\end{figure}

\subsection{Three-dimensional Trackings}
\begin{figure}
\centering
\includegraphics[width=3.5in]{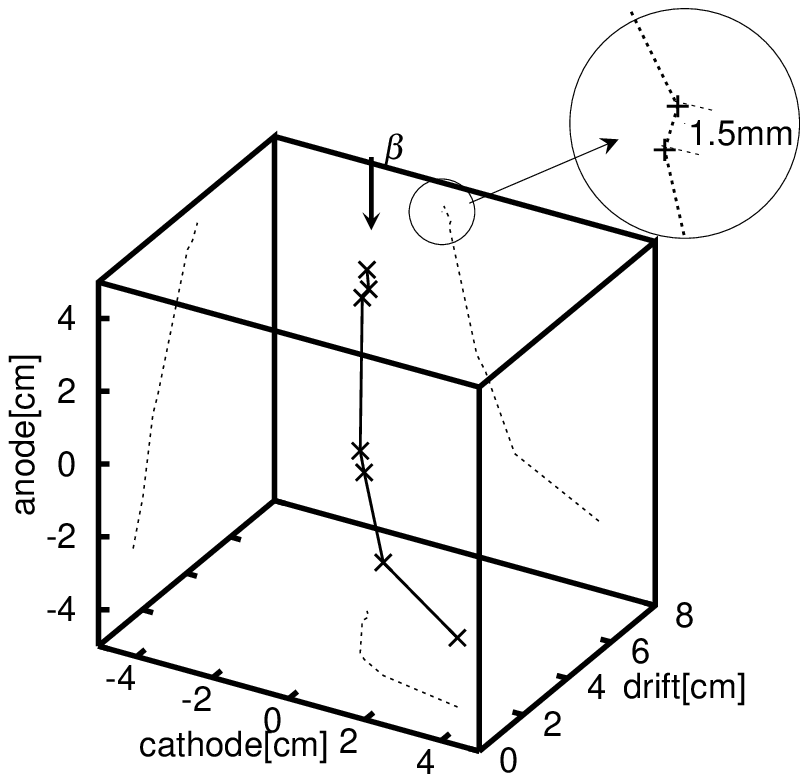}
\includegraphics[width=3.0in]{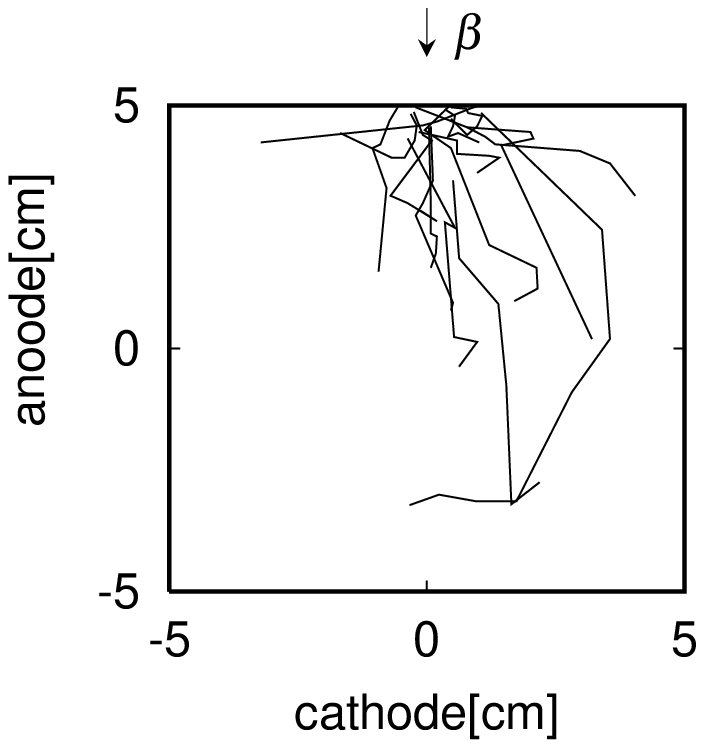}
\caption{A typical three-dimensional track of an electron (upper) and several tracks projected to the $\mu$-pic plane (lower).}
\label{3D_beta}
\end{figure}

We detected the tracks of the electrons from a $\rm {}^{90}Sr/{}^{90}Y$ radioactive source which emits electrons with the maximum kinetic energy of 0.546 MeV and 2.282 MeV. The Micro-TPC was triggered by a plastic scintillator of 6 $\times$ 6 $\times$ 2 mm${}^{3}$ placed between the radioactive source and the micro-TPC. The $\mu$-PIC was operated at the gas gain of 7000.
% with the gas mixture of argon 80 $\%$ and ethane 20 $\%$ at 1 atm. 
One of the typical three-dimensional tracks is shown in the upper panel of Fig.~\ref{3D_beta}, while the projections of the several tracks on the $\mu$-PIC plane are shown in the lower. We only detected the points where large energy deposition took place because of the insufficient gas gain, therefore the electron tracks were detected as the dispersed points. Since the principle of the micro-TPC is confirmed,  dense tracks like ones taken by a cloud chamber will be taken when we've achieved a sufficient gain.

%It is seen track structure as fine as 1 mm is detected by the micro-TPC, although we need to increase the gas gain to detect more precise tracks.
%It is seen that electron tracks are taken $$

\subsection{Development of the Prototype MeV Gamma-Ray Imaging Detector}
One of the most interesting applications of the micro-TPC is a MeV gamma-ray imaging detector which makes an event-by-event reconstruction possible\cite{MeVgamma_PSD:Orito}. Typical MeV gamma-ray imaging detectors of these days with double or multiple Compton method\cite{Comptel, MEGA}  measure the energy deposited to the electron ($E_{\rm e}$), the energy of the scattered gamma-ray ($E_{\gamma '}$) and the direction of the scattered gamma-ray ($\vec{{\gamma}'}$). The scattering angle of the gamma-ray ($\phi$) is calculated by the following equations.

\begin{eqnarray}
cos\phi&=&1-m_{e}c^2(1/E_{\gamma '}-1/E_{\gamma})\\
E_{\gamma}&=&E_{\gamma '}+E_e
\end{eqnarray}

\begin{figure}[h]
\centering
\includegraphics[width=2.in]{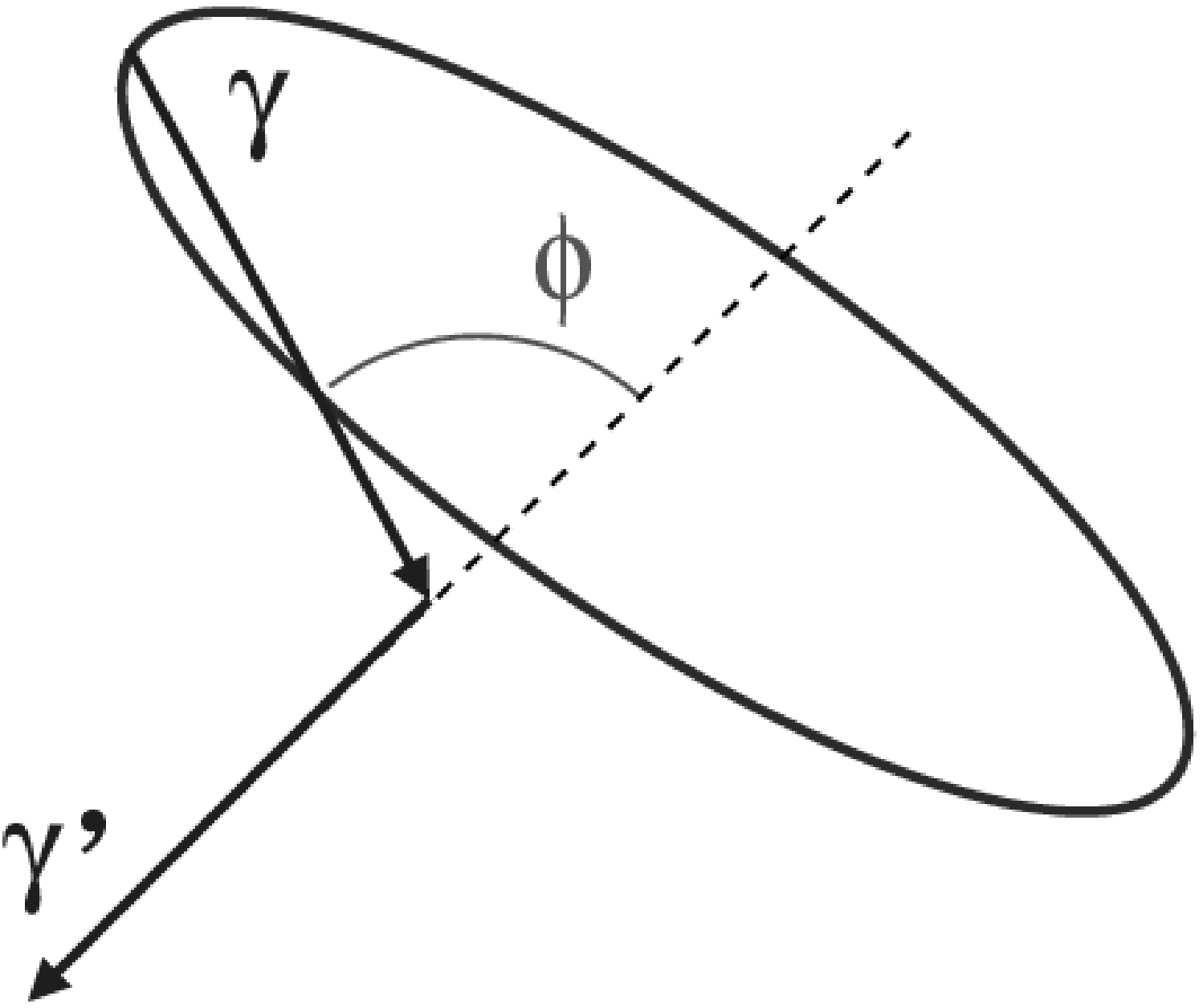}
\caption{Principle of the gamma-ray reconstruction by the double Compton method.}
\label{eventcone}
\end{figure}

Here, $m_e$ is the mass of the electron. The event cone is determined from $\vec{{\gamma}'}$ and $\phi$. The principle of the gamma-ray reconstruction with double Comptom method is shown in Fig.  \ref{eventcone}. 

\begin{figure}[h]
\centering
\includegraphics[width=2.in]{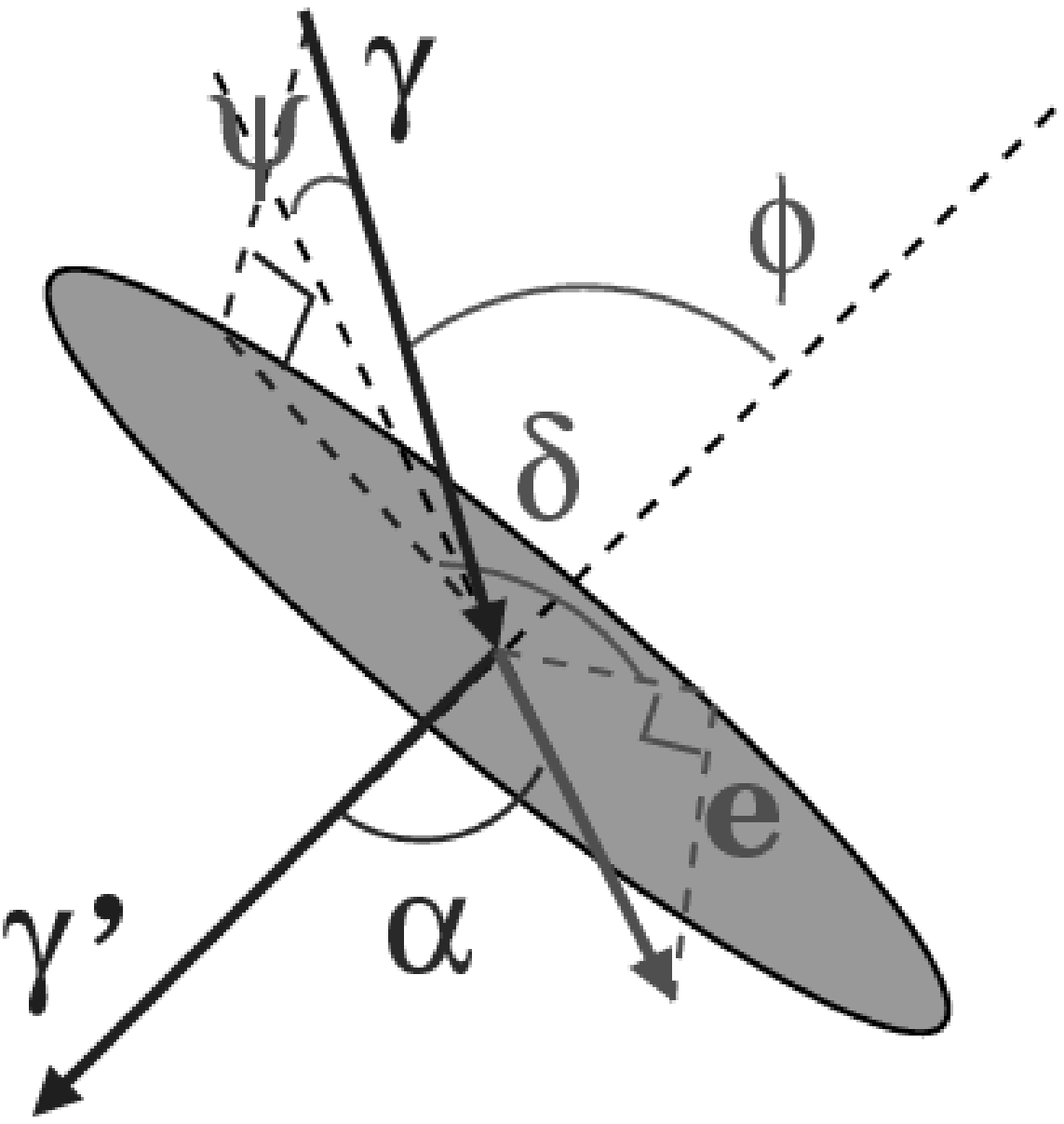}
\caption{Principle of the gamma-ray reconstruction by this new concept detector.}
\label{MeV_prin}
\end{figure}

 Our new concept for the MeV gamma-ray imaging is to measure the direction of the Compton scattered electrons ($\vec{e}$) with the micro-TPC in addition to $E_{\rm e}$, $E_{\gamma '}$, and $\vec{{\gamma}'}$. Since the micro-TPC can detect the track of the electrons as we have shown in the previous section, the hybrid detector of the micro-TPC and the enclosing scintillators can realize this new concept. With this new concept detector, we can determine another angle $\delta$ so that the incident gamma-ray is reconstructed for each event.
%event-by-event reconstruction of the incident gamma-ray is realized.  
The principle is shown in Fig. \ref{MeV_prin}. 

We have a redundant measured value $\alpha$ which can be used for the background rejection.  Since no collimators are necessary, a large field of view can be realized.
%We can also measure the geometrical angle $\alpha$ and use it for the background reduction since $\alpha$ is not used for the reconstruction.  Since no collimators is necessary, a large field of view can be realized.

\begin{figure}
\centering
\includegraphics[width=3.5in]{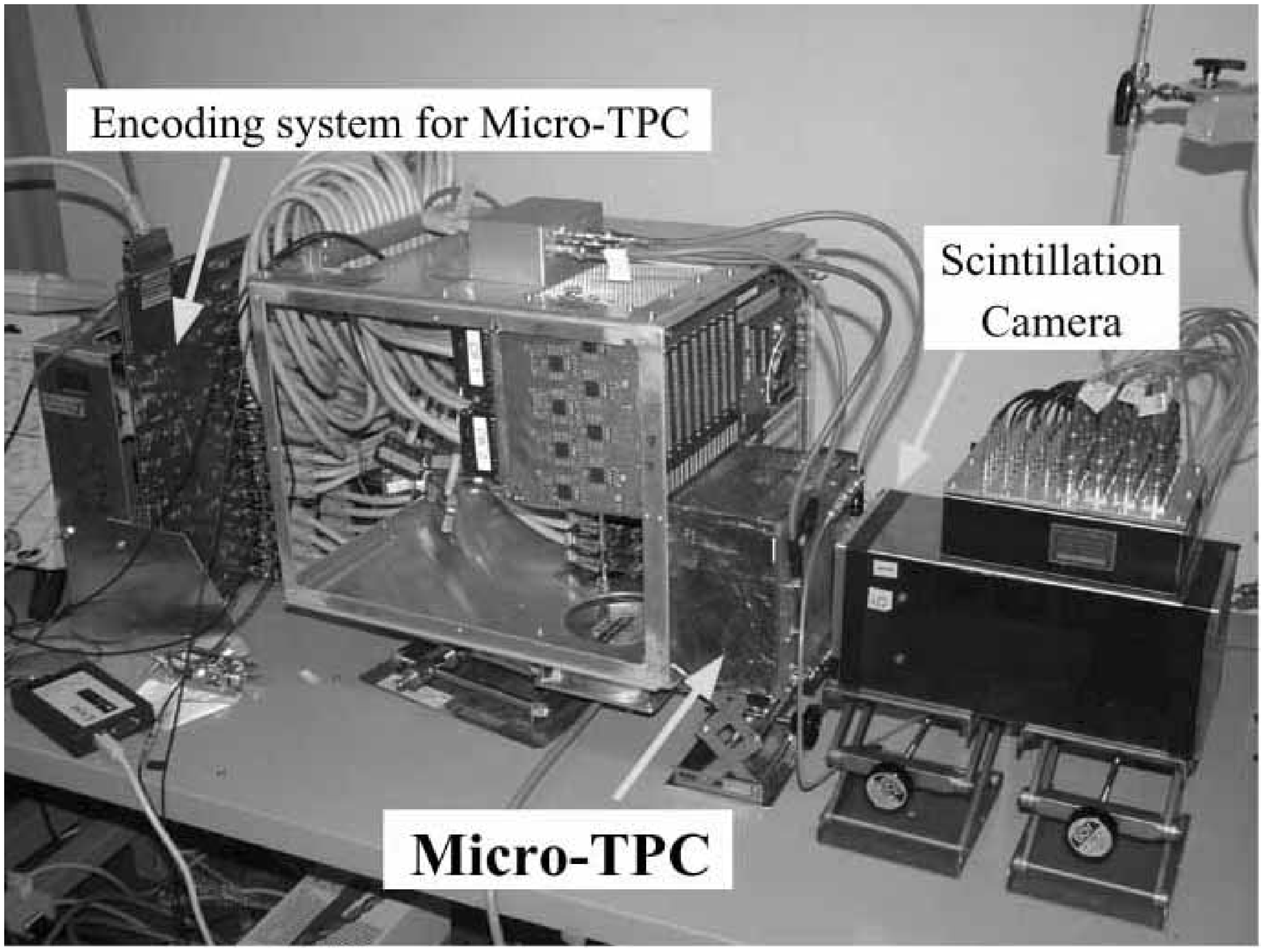}
\caption{The prototype of the MeV gamma-ray imaging detector.}
\label{prototype_photo}
\end{figure}

We developed a prototype of the MeV gamma-ray imaging detector with the micro-TPC of 10 $\times$ 10 $\times$ 8 $\rm cm^3$ volume and an NaI (Tl) scintillator of 4''$\times$4''$\times$1'' size read by 25 single anode PMTs of 3/4'' diameters (Hamamatsu~R116). The picture of the prototype is shown in Fig. \ref{prototype_photo}.

\begin{figure}
\centering
\includegraphics[width=4.0in]{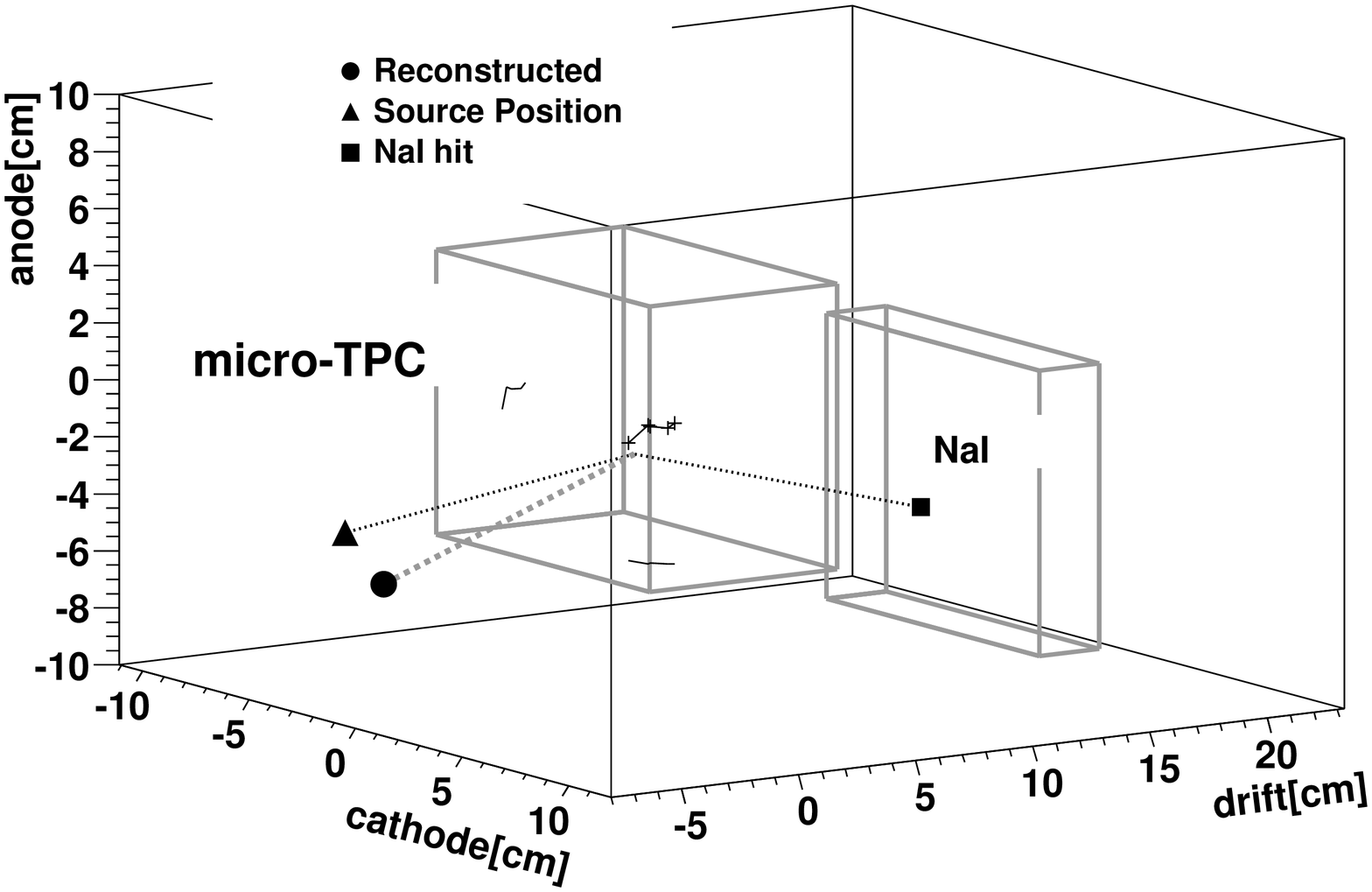}
\caption{One of the well-reconstructed gamma-rays taken by the prototype MeV gamma-ray detector.}
\label{reconstructed}
\end{figure}

We tested the prototype with a radioactive source of $\rm {}^{133}$Ba which mainly radiates 356~keV gamma-rays. The setup is shown in Fig. \ref{reconstructed}. We set the origin of the coordinates at the center of the $\mu$-PIC detection plane and X, Y and Z axes are taken along the cathode, anode and drift direction of the micro-TPC, respectively. The radioactive source was set at (-4.0,-4.3,-4.8) without any collimators and the center of the NaI (Tl) scintillator was set at (-5.5,-5.5,22.9).

Electron tracks are detected by the micro-TPC, while the energy and the position of the gamma-rays are taken by the NaI (Tl) scintillator. A gate of two-$\mu$s was opened after the NaI (Tl) trigger for the micro-TPC signals and coincident events were recorded. We operated the $\mu$-PIC with a gas gain of 5000.
% flowing a gas mixture of argon 80 $\%$ and ethane 20 $\%$ at 1atm.
For this concept test, we did not use the information on the electron energy but assumed $E_e=E_{\gamma}-E_{\gamma '}$ as we knew the source energy.  This test has much meaning since the MeV gamma-ray imaging for the radioactive sources of a know energy is useful for the medical use. 

One of the well-reconstructed events is shown in Fig. \ref{reconstructed}. The source is shown by the closed triangle. The electron track is shown in the solid line and its projections are also drawn by the thin solid lines. The measured energy by the NaI (Tl) scintillator was 260 keV and the calculated position of the gamma-ray in the  NaI (Tl) scintillator is shown by the closed square.  We reconstructed the direction of the incident photon from $E_{\gamma}$, $E_{\gamma '}$, $\vec{{\gamma}'}$ and $\vec{e}$. The reconstructed point in the Z=-4.8 plane is (-2.2, -5.7, -4.8) and is shown by the filled circle. The reconstruction was realized by the angular accuracy of  $10^{\circ}$ for this event. The gamma-ray is well-reconstructed in this event so that we can claim that this new concept is a very promising method for the MeV gamma-ray imaging.

We are going to use the energy information of the electrons, so that we can realize the MeV gamma-ray imaging for the gamma-rays of unknown energies, or we can use the energy for the background rejection. Statistical studies are also under way and the angular resolution and the detection efficiency will appear in another report.

%We need to increase the gas gain to achieve higher detection efficiency and more accurate reconstructions.  Then we can measure the 

% Studies on the efficiency and accuracy of the reconstruction will be performed in the near future.

\section{Conclusion}
We improved the $\mu$-PIC with the detection area of $\rm 10 \times 10 cm^2$ and obtained the maximum gas gain of more than $\rm 10^4$ without any other multipliers. We detected X-rays of the rate as high as 7.7 MHz with the $\mu$-PIC and the pipe-line readout electronics. We developed a micro-TPC with the $\mu$-PIC readout and took three-dimensional electron tracks. We developed a prototype of the MeV gamma-ray imaging detector with the micro-TPC and an NaI (Tl) scintillator so that we showed that this is a promising detector for the MeV gamma-ray imaging.

% or

% Use this command to get the appendices' numbers in "A", "B" instead of the
% default capitalized Roman numerals ("I", "II", etc.).
% However, the capital letter form may result in awkward subsection numbers
% (such as "A-A"). Capitalized Roman numerals are the default.
%\useRomanappendicesfalse
%
%\appendices

%\section{}

\section*{Acknowledgment}
% optional entry into table of contents (if used)
%\addcontentsline{toc}{section}{Acknowledgment}
This work is supported by a Grant-in-Aid in Scientific Research of the Japan Ministry of Education, Culture, Science, Sports and Technology, and ``Ground Research Announcement for Space Utilization'' promoted by Japan Space Forum.



% http://www.ctan.org/tex-archive/macros/latex/contrib/supported/IEEEtran/testflow/bibtex
%\bibliographystyle{IEEEtran.bst}

% insert where needed to balance the two columns on the last page
%\newpage

%\vfill


\end{document}